\documentstyle[]{mn}

\footnotesize
\newdimen\minuswidth    
\setbox0=\hbox{$-$}
\minuswidth=\wd0
\catcode`@=\active
\def@{\kern\minuswidth}
\newdimen\digitwidth    
\setbox0=\hbox{\rm0}
\digitwidth=\wd0
\catcode`!=\active
\def!{\kern\digitwidth}
\normalsize

\title[PSR J1811$-$1736] {The Parkes Multibeam Pulsar Survey: 
PSR~J1811$-$1736 -- a pulsar in a highly eccentric binary system}
\author[A. G. Lyne et al.]
{A. G. Lyne$^1$,
F. Camilo$^1$,
R. N. Manchester$^2$, 
J. F. Bell$^2$, 
V. M. Kaspi$^3$,
\newauthor
N. D'Amico$^{4,5}$, 
N. P. F. McKay$^1$,
F. Crawford$^3$,
D. J. Morris$^1$,
\newauthor
D. C. Sheppard$^1$,
I. H. Stairs$^1$ \\
$^1$ University of Manchester,
Jodrell Bank Observatory, Macclesfield, Cheshire, SK11~9DL, UK\\
$^2$ Australia Telescope National Facility, CSIRO, P.O.~Box~76, Epping
NSW~1710, Australia\\
$^3$ Massachusetts Institute of Technology, Center
for Space Research, 70 Vassar Street, Cambridge, MA~02139, USA\\
$^4$ Osservatorio Astronomico di Bologna, via Ranzani 1, 40127
Bologna, Italy\\
$^5$ Istituto di Radioastronomia del CNR, via Gobetti 101, 40129
Bologna, Italy\\
}

\date{1999 November 16}
\begin{document}

\maketitle
\newcommand{\setthebls}{
}

\setthebls

\begin{abstract} 
We are undertaking a high-frequency survey of the
Galactic plane for radio pulsars, using the 13-element multibeam
receiver on the 64-m Parkes radio telescope.  We describe briefly the
survey system and some of the initial results. PSR~J1811$-$1736, one
of the first pulsars discovered with this system, has a rotation
period of 104\,ms.  Subsequent timing observations using the 76-m
radio telescope at Jodrell Bank show that it is in an 18.8-day,
highly-eccentric binary orbit.  We have measured the rate of advance
of periastron which indicates a total system mass of $2.6\pm
0.9$\,M$_{\odot}$, and the minimum companion mass is about
0.7\,M$_\odot$.  This, the high orbital eccentricity and the recycled
nature of the pulsar suggests that this system is composed of two
neutron stars, only the fourth or fifth such system known in the disk
of the Galaxy.  
\end{abstract}


\section{INTRODUCTION}

A multibeam receiver operating at a wavelength of 20\,cm has
recently been installed on the 64-m Parkes radio telescope in
Australia. This receiver permits simultaneous observation of 13
regions of sky, increasing the speed of surveys by the same factor.
While the main motivation for the development of this system was to
survey the sky for H{\sc i} in the local universe \cite{swb+96}, it
also enables very efficient surveys for pulsars. The central observing
frequency of 1374\,MHz is higher than usual for all-sky pulsar
surveys, but is ideal for surveys at low Galactic latitudes. At lower
frequencies, sensitive surveys of the Galactic plane are limited by
the high background temperature, as well as broadening of the pulses
caused by dispersion and scattering of the pulsed emission by the
interstellar electron plasma.  All of these effects are strongly
frequency-dependent and are much reduced at 1374\,MHz. The Jodrell
Bank and Parkes surveys of Clifton et al.~(1992)\nocite{clj+92} and
Johnston et al.~(1992)\nocite{jlm+92} were centred close to this
frequency and both were particularly successful at finding many
distant and young pulsars, several of which are also X-ray and/or
gamma-ray emitters and some possibly associated with supernova
remnants.

We are undertaking a survey for pulsars in an $8\degr$-wide strip
along the Galactic plane using the multibeam receiver on the Parkes
telescope. This survey is proving remarkably successful and the
discovery of the first $\sim 100$ pulsars is presented elsewhere
(Manchester et al., in preparation).
We briefly describe the experimental details of the
survey in Section~2.  Detailed timing observations to determine
accurate pulsar positions, pulse periods, period derivatives and other
parameters are an essential follow-up to any pulsar survey. Timing
observations and analysis procedures for the multibeam survey pulsars
are described in Section~3.  

Finally, Section~4 describes a particularly interesting pulsar,
PSR~J1811$-$1736. This pulsar, one of the first discovered in the
survey, is in a very eccentric orbit with a companion which is
probably another neutron star.  Pulsars in such binary systems in the
Galactic plane are rare, there being only three known which are definite
(PSRs~J1518+4904, B1534+12 and B1913+16) and one which is possible
(PSR~B1820$-$11).  Such systems are of great interest because both
stars act essentially as point masses and they make excellent systems
for studying their dynamics with high precision (e.g. Damour \& Taylor
1992)\nocite{dt92}.  In particular, the discovery of more systems like
these will allow additional sensitive tests of general relativity and
alternative theories of gravity, more precise measurements of neutron
star masses, and better estimates of the rate of coalescence of
neutron-star binaries, events which are likely to be detected by
future gravitational-wave detectors.

\section{THE PARKES MULTIBEAM SURVEY}

The survey of the Galactic plane for pulsars using the multibeam
receiver on the Parkes radio telescope commenced in 1997 August.  The
survey observations are made in a 288-MHz band centred on 1374\,MHz.
Receivers for each of the 13 beams are dual-channel cryogenic systems
sensitive to orthogonal linear polarisations. The full width at
half-power of each beam is approximately 14~arcmin and, at high
Galactic latitudes, the system noise temperature of each channel is
about 21\,K. A large filterbank system was constructed to process the
data from the 13 beams.  The filterbank has 96 3-MHz channels covering
the 288-MHz bandwidth for each polarisation of each beam, making a
total of 2496 channels. Detected signals from individual frequency
channels are added in polarisation pairs, high-pass filtered with a
cutoff at approximately 0.2\,Hz, integrated and 1-bit digitised every
250\,$\mu$s, and recorded on magnetic tape (DLT) for subsequent
analysis.

It is planned that the survey will cover the region $\pm4\degr$ about
the Galactic plane from longitude $260\degr$ to $50\degr$.  The
relatively long observation time of 35\,min per pointing, combined
with the excellent receiver noise performance and the wide bandwidth,
gives a high sensitivity, with a limiting flux density for long-period
pulsars of about 0.15\,mJy. This is about seven times better than the
previously most sensitive surveys of the Galactic plane (Clifton et
al.~1992; Johnston et al.~1992).  Survey observations are made on a
hexagonal grid of pointings to give complete sky coverage, with
adjacent beams overlapping at the half-power points.  A total of 2136
pointings of the 13 beams are required to cover the survey region. At
the time of writing, approximately 60 per cent of these pointings have
been observed.

Analysis procedures are similar to those used in previous surveys
(e.g. Manchester et al.~1996)\nocite{mld+96} with de-dispersion
followed by transformation into the modulation frequency domain using
a Fast Fourier Transform (FFT) and harmonic summing to improve
sensitivity to the typically narrow pulses.  Some periodic
interference is present in the data and this is removed primarily by
ignoring signals found repeatedly in undedispersed power spectra, both
in simultaneous data from different beams as well as from single beams
at many different times.  Processing is currently being carried out on
networks of workstations.  Initially, a
single coherent FFT has been performed over each 35-min data set,
giving reduced sensitivity to pulsars in short-period binary systems
in which the period is varying due to the Doppler effects of the pulsar
motion. The data are now being reprocessed in order to search for such
accelerated signals.

Pulsar candidates from the initial processing are scheduled for
reobservation in order to confirm their identity as pulsars. We use
the centre beam of the multibeam system to make five 6-minute
observations, at the nominal position and four surrounding points.  In
most cases, this gives detections at two or three positions, confirming
the pulsar and allowing an improved position to be determined with an
accuracy of about 2 arcmin.  Where no detection results from these
observations, confirmation is attempted by means of a 35-min
observation made at the nominal position.  The improved positions
usually permit much shorter observation times in the follow-up timing
measurements, because of the higher signal-to-noise ratio obtained
when the pulsar is in the centre of the telescope beam.  Furthermore,
they allow a smaller range of parameter space to be searched in
obtaining coherent timing solutions, reducing the overall number of
observations required.

At the time of writing, about 90 per cent of the observed pointings
have been processed using the ``non-acceleration'' search code, whilst
analysis with the ``acceleration'' search code has just started.
Confirmation observations have been made on most of the better
candidates, resulting in the detection of about 400 previously unknown
pulsars, most of which are relatively distant, the median distance
being $\sim7$~kpc.  More than 150 previously
known pulsars have also been detected. The current discovery rate is
an unprecedented one pulsar per hour of survey observing time.
This rate will decline as regions further from the Galactic
plane are surveyed, but we estimate the survey will detect at least
600 previously unknown pulsars, nearly doubling the number known
before the survey commenced, and providing a significant database for
many different follow-up studies.

\section{TIMING OBSERVATIONS AND ANALYSIS}

After confirmation, each pulsar is subjected to a series of timing
observations at either the Parkes 64-m telescope or the
Lovell 76-m telescope at Jodrell Bank.  Almost all of the detected
pulsars north of declination $-35\degr$ are being timed at Jodrell
Bank. At Parkes, data are recorded using just the central beam of the
multibeam system. The Jodrell Bank observations are made in a 96-MHz
band centred on 1376\,MHz.  Dual-channel cryogenic systems receiving
orthogonal circular polarisations are used, each channel having a
system noise temperature of about 30\,K at high Galactic latitude.
Each of the two polarisation signals are down-converted and fed into a
multi-channel filterbank consisting of 32~3-MHz filters, before
digitisation.

Data from both Parkes and Jodrell Bank are de-dispersed and
synchronously folded at the topocentric pulsar period for $\sim1$\,min
to form sub-integrations; the total integration time per observation
ranges between 2 and 30\,min, dependent upon the pulsar flux
density.  Each pulse profile obtained by summing over an observation is
convolved with a high signal-to-noise ratio ``standard profile'' for the
corresponding pulsar, producing a topocentric time-of-arrival (TOA).
These are then processed using the {\sc tempo} program (see
http://pulsar.princeton.edu/tempo) which converts them to barycentric
TOAs at infinite frequency and performs a multi-parameter fit for the
pulsar parameters.  Barycentric corrections are obtained using the Jet
Propulsion Laboratory DE200 solar-system ephemeris \cite{sta82}. Except
for especially interesting cases, we make timing observations of each
pulsar over about 12 months, resulting in an accurate position, period,
period derivative and dispersion measure (DM).  This provides the basic
parameters necessary for follow-up studies such as investigations of
the Galactic distribution of pulsars, studies of the interstellar
medium and high-energy observations.

These observations also reveal pulsars which are members
of binary systems. After detection, such pulsars are observed more
intensively to determine their binary parameters. Of the new pulsars
detected so far, seven have already been identified as members of
binary systems and one of these is described in the next section.

\begin{figure*}
\setlength{\unitlength}{1cm}
\begin{picture}(12,10)
\put(-2,11){\includegraphics{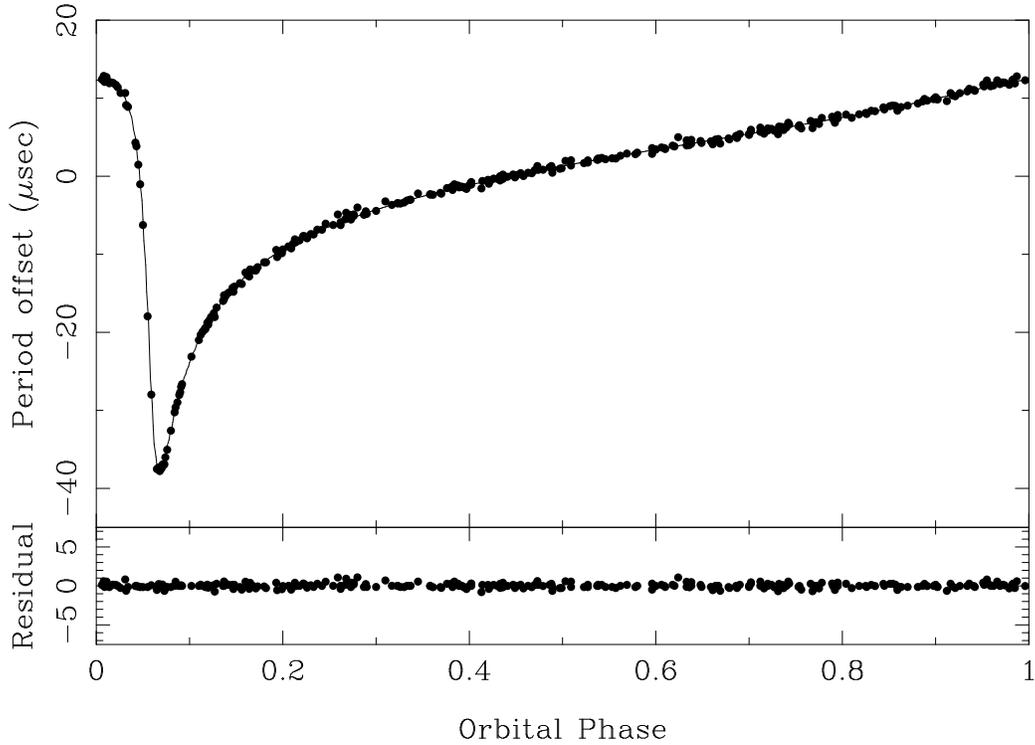}}
\end{picture}
\caption
[] { The observed variation of the barycentric period of
PSR~J1811$-$1736 over the 18.8-day orbital period measured using the 76-m
Lovell telescope at Jodrell Bank (top).  The residuals from the
model given in Table 1 and shown as the continuous line in
the top diagram are shown in the bottom plot. The orbital phase is 
measured from the time of the ascending node.}
\label{fig:perphase}
\end{figure*}

\section{PSR~J1811$-$1736} 

PSR~J1811$-$1736 was initially observed on 1997 August 3 and revealed
as a candidate with a 104-millisecond period and DM of
477\,cm$^{-3}$\,pc. It was confirmed as a pulsar on 1997 December 19.
This observation gave a period which was substantially different from
the initial discovery period, indicating a possible binary
nature. This was confirmed by a series of observations at Jodrell Bank
during the beginning of 1998 which demonstrated that the pulsar was a
member of a binary system in a highly eccentric orbit.
Fig.~\ref{fig:perphase} shows the measured period variation of
PSR~J1811$-$1736 through the orbital period of 18.8 days.
Subsequently, we performed a phase-coherent analysis of the arrival
times over a 16-month period.  The results are summarised in Table~1.
Errors given in parentheses refer to the last quoted digit and are
twice the formal standard error.

Using the Taylor \& Cordes~(1993)\nocite{tc93} model for the electron
density distribution in the Galaxy and the measured DM, we obtain an
estimated distance for PSR~J1811$-$1736 of $\sim6$\,kpc.  With the
measured flux density at 1400\,MHz of $S_{1400} = 0.7$\,mJy (Table~1),
the luminosity of this pulsar at 1400\,MHz is $L_{1400}\equiv S_{1400}
d^2 = 25$\,mJy\,kpc$^2$.  The pulsar has not yet been detected at
other frequencies, mainly because of the large amount of scattering at
lower frequencies. However, we can get a rough estimate of its
luminosity at $\sim 400\,$MHz by assuming a typical pulsar spectral
index of $\sim -1.6$.  We obtain $L_{400}\approx200$\,mJy\,kpc$^2$,
indicating that this is a relatively luminous pulsar, significantly
above any low-luminosity cut-off in the luminosity distribution of
Galactic disk pulsars at 400\,MHz of 1\,mJy\,kpc$^2$ or lower (Lorimer
et al.~1993; Lyne et al.~1998).\nocite{lbdh93,lml+98}

\begin{table}
\begin{center} 
\begin{tabular}{ll} 
\multicolumn{2}{c}{Table 1. Parameters of PSR~J1811$-$1736}                            \\
\hline 
\hline 
\multicolumn{2}{c}{Measured Parameters}                                       \\
\hline
Right Ascension (J2000)                 & $18^{\rm h}11^{\rm m}55\fs01(1)$    \\
Declination (J2000)                     & $-17^{\circ}36\arcmin 36\farcs9(13)$\\
Dispersion Measure (cm$^{-3}$ pc)       & 477(10)                              \\
Period (s)                              & 0.104181954734(3)                   \\
Epoch of Period (MJD)                   & 51050.0                             \\
Period Derivative                       & $1.8(6)\times 10^{-18}$             \\
Orbital Period (days)                   & 18.779168(4)                        \\
Projected Semi-major Axis (lt-s)        & 34.7830(8)                          \\
Eccentricity                            & 0.82802(2)                          \\
Epoch of Periastron (MJD)               & 51044.03702(3)                      \\
Angle of Periastron (degrees)           & 127.661(2)                          \\
Rate of Advance of Periastron ($\degr$\,yr$^{-1}$) & 0.009(2)                 \\
Rate of Change of Projected Semi-major Axis & $<3\times 10^{-11}$             \\
R.M.S. Timing Residual (ms)             & 1.0                                 \\
Flux Density at 1400 MHz (mJy)          & 0.7(2)                              \\
Interstellar Scattering at 1376\,MHz (ms)& 21(3)                              \\
\hline
\multicolumn{2}{c}{Derived Parameters}                                        \\
\hline
Galactic Longitude (degrees)            & 13.4                                \\
Galactic Latitude (degrees)             & 0.7                                 \\
Distance (kpc)                          & $6(1)$                              \\
Characteristic Age (yr)                 & $9_{-2}^{+4}\times 10^8$            \\
Surface Magnetic Field (G)              & $1.4(2)\times 10^{10}$              \\
Total System Mass (M$_{\odot}$)         & 2.6(9)                              \\
Mass Function (M$_{\odot}$)             & 0.128                               \\
\hline 
\end{tabular} 
\end{center} 
\end{table}

\begin{figure*}
\setlength{\unitlength}{1cm}
\begin{picture}(12,10)
\put(-3,11){\includegraphics{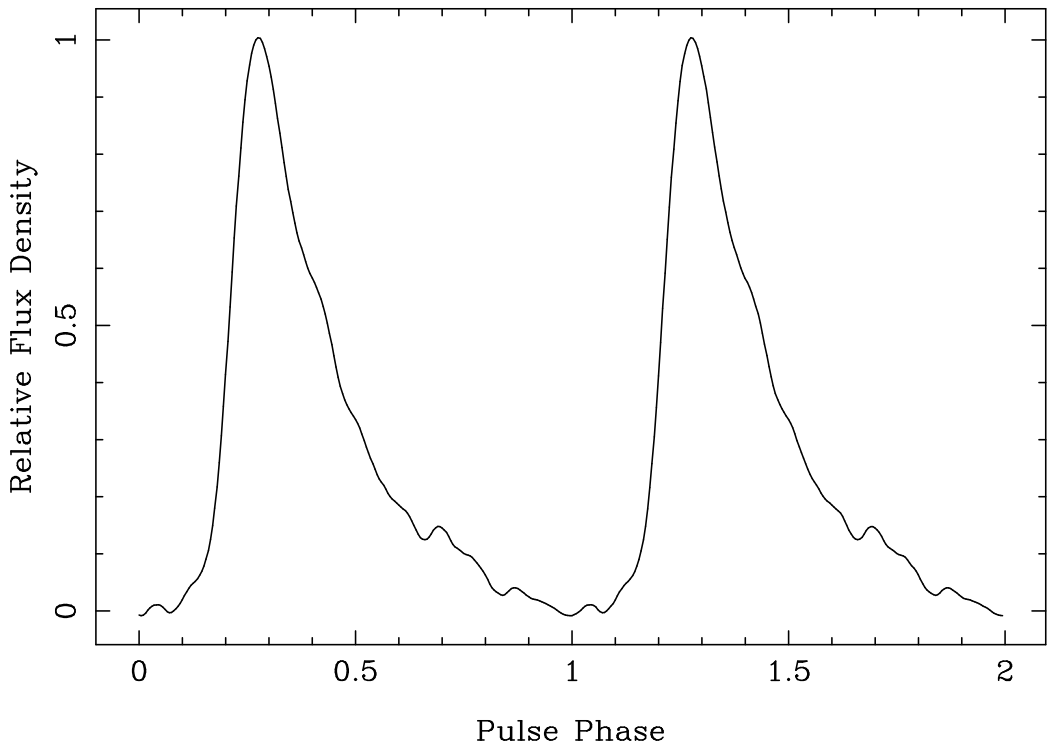}}
\end{picture}
\caption
[] { The pulse profile of PSR~J1811$-$1736 observed at a central
frequency of 1376\,MHz, covering two
complete periods of pulse phase.  The time resolution is approximately
4 milliseconds or 0.04 in pulse phase.  The features on the trailing
edge are at about the level expected from random noise and are
unlikely to be significant.}
\label{fig:profile}
\end{figure*}

The average pulse profile of PSR~J1811$-$1736 is shown in
Fig.~\ref{fig:profile}.  It is very asymmetric, with a rapidly rising
leading edge with an approximately exponential tail following the
peak.  The tail is detectable throughout the period and is
characteristic of the effects of multi-path propagation in the
interstellar medium, more usually observed at lower frequencies.  In
such scattering, the timescale of the exponential, $\tau_s$,
varies with frequency, $\nu$, approximately as $\tau_s \propto
\nu^{-4}$.  Over the 288-MHz band of the Parkes receiver, $\tau_s$ is
found to vary by about a factor of two, consistent with such a
variation.  The amount of scattering is unusually large for a pulsar
with this DM observed at around 1400\,MHz: the Taylor \&
Cordes~(1993)\nocite{tc93} model predicts $\tau_s \approx 2$\,ms, while
the observed $\tau_s$ is about 20\,ms.  This accounts for the rather
large rms timing residual reported in Table~1.

As listed in Table~1, the spin parameters for PSR~J1811$-$1736 imply a
large ``characteristic age'' of $\tau = P/2\dot P =
9\times10^8$\,yr, and a relatively low implied surface dipole
magnetic field strength of $B = 3.2\times10^{19}\sqrt{P \dot P} =
1.4\times10^{10}$\,G.  Such spin parameters are typical of
the known double-neutron star systems (Taylor et al.~1995; Nice,
Sayer, \& Taylor~1996\nocite{tmlc95,nst96}). This pulsar is
located below the ``spin-up line'' on a period--magnetic field diagram
(see, e.g. Lyne \& Graham-Smith, 1998, p115 or Phinney and Kulkarni
1994, Fig. 1)\nocite{ls98,pk94}, suggesting that
it was spun-up at an earlier stage in its history.

\begin{figure*}
\setlength{\unitlength}{1cm} 
\begin{picture}(12,10)
\put(0.5,0){\includegraphics{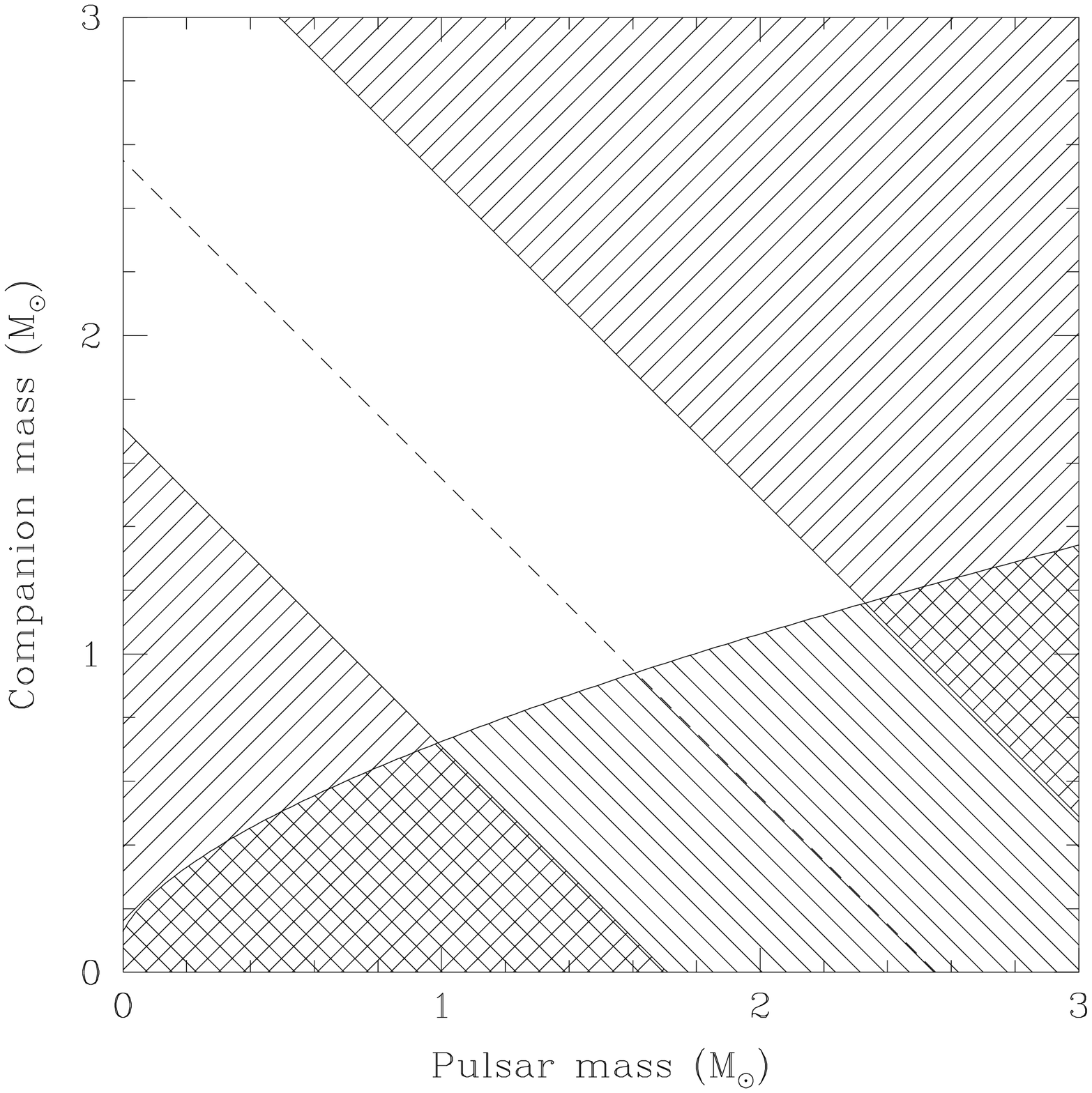}} 
\end{picture} 
\caption [] { Constraints on the masses of PSR~J1811$-$1736 and its
companion, obtained from the observed mass function and the measured
rate of advance of periastron.  The allowed component masses lie in the
clear area of the diagram (see text).}
\label{fig:m1m2} 
\end{figure*}

In addition to the standard astrometric, spin, and Keplerian orbital
parameters, a satisfactory fit to the timing data requires inclusion
of the rate of orbital precession of the pulsar in the fitted
parameters, giving $\dot\omega = 0.009\pm0.002 \degr$\,yr$^{-1}$.
Assuming that the companion star is a compact object, that tidal effects
are negligible and that the measured rate of advance of periastron is due
to general relativistic effects alone, we have (e.g. Will
1993):\nocite{wil93}
\begin{equation}
\dot \omega = 3(2\pi/P_b)^{5/3}(1-e^2)^{-1}T_{\odot}^{2/3}(m_1+m_2)^{2/3},
\end{equation}
where $T_{\odot} = G M_{\odot}/c^3 = 4.925\times10^{-6}$\,s, $P_b$ is
the binary period, $e$ the binary eccentricity, and $m_1$ and $m_2$ are
the pulsar and companion masses, in solar masses.  The
measurement of $\dot \omega$ yields a total system mass of $m_1 + m_2 =
2.6\pm0.9$ M$_\odot$, or an average component mass of
$1.3\pm0.4$\,M$_\odot$.  This compares with the average mass of
neutron stars measured in other binary pulsar systems of
$1.35\pm 0.04$\,M$_\odot$ \cite{tc99}.
The constraints on the masses of the two stars
can be most readily seen in Fig.~\ref{fig:m1m2}.  The region under
the convex curve is excluded because of the requirement that $\sin i
\le1$ in the ``mass function'',
\begin{equation}
f_1(m_1, m_2, i) = \frac{(m_2 \sin i)^3}{(m_1+m_2)^2} = 
 \frac{4 \pi^2}{T_{\odot}} \frac{(a_1 \sin i)^3}{P_b^2},
\end{equation}
where $a_1 \sin i$ is the projected semi-major axis of the pulsar
orbit.  From Fig.~\ref{fig:m1m2} we see that $m_1 < 2.3$\,M${_\odot}$
and that $m_2 > 0.7$\,M${_\odot}$.  These constraints on the
individual and average masses of the system, the lack of eclipses, the
low magnetic field of the pulsar, and the high binary eccentricity, all
suggest that the pulsar companion is another neutron star.  If so,
this is only the fourth or fifth double-neutron star system discovered
in the Galactic disk (e.g. Nice, Sayer \& Taylor~1996)\nocite{nst96}.

If the companion were a main sequence or giant star, in addition to
relativistic effects, classical effects could also contribute to the
measured value of $\dot\omega$ (e.g. Lai, Bildsten \& Kaspi 1995;
Kaspi et al. 1996; Wex 1998)\nocite{lbk95,kbm+96,wex98}.  For a pulsar of mass
$m_1=1.35$\,M${_\odot}$, the relativistic contribution always
dominates.  However, for a small range of companion masses around
1\,M${_\odot}$, some remaining $\dot\omega$ could in principle be
attributed to a classical quadrupole moment $Q$ in the companion
star.  The neutron star should raise a tide in a companion having
radius $R_2$ such that $Q \sim k m_1 R_2^2 (R_2 / r)^3,$ where $k$ is
the apsidal constant and $r$ is the instantaneous centre-of-mass
separation \cite{lbk95}.  However, given the orbital parameters and
estimates for $k$ from stellar modeling
\cite{cg91,hej87}, the values of $Q$ permitted by the observations are
much smaller than would be expected.  Thus, a main-sequence or giant
companion should result in a much larger $\dot\omega$ than is
observed.  This conclusion is not very sensitive to the choice of
$m_1$.

We note that a rapidly rotating main sequence star companion, with
spin axis sufficiently misaligned with the orbital angular momentum,
could result in a spin-induced quadrupole that precesses the orbit in
the retrograde direction, cancelling the tidal effect.  However, this
is most unlikely as it requires fine-tuning the spin and tidal effects
to be surprisingly close in magnitude.

Recently, the companion in another double neutron-star candidate
system (PSR~B2303+46) has been shown to be a white dwarf
\cite{vk99}. It is therefore important to assess what other
constraints exist on the nature of the companion.  Searches of a
number of optical telescope archives, revealed that the only available
images were from the Digital Sky Survey V band, obtained at the
UK Schmidt telescope. No sources are visible in the vicinity of the
pulsar down to the plate limit of M$_{V} \sim 16$. Unfortunately, this
can only conclusively rule out a red giant companion and a deeper
image would be required to test if the companion is a main sequence
star (M$_{V} \sim 19$) or a white dwarf (M$_{V} \sim 27$).

We note that there is no evidence of occultation of the pulsar at any
phase of the orbit (Fig. 1). The closest approach of the two stars
depends upon the unknown inclination angle $i$, but is several solar
radii, making the occultation of the pulsar by a solar-mass
main-sequence companion star possible, but rather unlikely.  These
observations do not therefore distinguish between a main-sequence
and a compact companion.  However, the whole orbit is smaller
than a giant, ruling out the possibility of such a companion.

The absence of significant tidal or quadrupole effects on $\dot
\omega$ suggests that the companion star is compact and therefore
either another neutron star or a white dwarf.  Such binary systems are
generally thought to evolve from two initially massive stars.  The
formation of a neutron star in the supernova collapse of the primary
is expected in due course to be followed by a common-envelope and
spiral-in stage.  During this time, it is detectable as an accreting
high-mass X-ray binary system (e.g. Verbunt~1993)\nocite{ver93}, prior
to the formation of a second neutron star or a white dwarf.  If the
pulsar is the second-formed compact object, it is not clear how the
magnetic field became so small, an attribute which is generally
thought to arise during mass-transfer.  If the pulsar is the original
neutron star, the ensuing mass-transfer would have reduced the
magnetic field.  However, this would also have circularised the orbit
which would have remained that way if the companion evolved to a white
dwarf.  Since the orbit is eccentric, we must conclude that it was
probably the formation of a second neutron star which gave rise to the
eccentricity.

In the future, this binary system is not likely to evolve
significantly.  Because of its relatively large orbital size, it does
not emit significant amounts of gravitational radiation: using the
formulae of Peters~(1964)\nocite{pet64}, we calculate that the
coalescence time of the PSR~J1811$-$1736 system is about
$10^{12}$\,yr.  Therefore, systems such as this make a negligible
contribution to the neutron-star merger-rate calculations which are
important for gravity-wave detectors such as LIGO.

\section*{Acknowledgements} 
We gratefully acknowledge the technical assistance provided by George
Loone, Tim Ikin, Mark Leach and all the staff at the Parkes
Observatory in developing the pulsar multibeam system.  The
Parkes radio telescope is part of the Australia Telescope which is
funded by the Commonwealth of Australia for operation as a National
Facility managed by CSIRO.  F. Camilo gratefully acknowledges support
from the European Commission through a Marie Curie fellowship under
contract no. ERB~FMBI~CT961700.  V. M. K. is an Alfred P. Sloan
Research Fellow.


\begin{thebibliography}{{Staveley-Smith {\rm et~al. }}{1996}}

\bibitem[\protect\citename{Claret \& Gimenez }{1991}]{cg91}
Claret~A., Gimenez~A., 1991, Astr. Astrophys. Suppl. Ser., 87, 507

\bibitem[\protect\citename{Clifton {\rm et~al. }}{1992}]{clj+92}
Clifton~T.~R., Lyne~A.~G., Jones~A.~W., McKenna~J., Ashworth~M., 1992, Mon.
  Not. R. astr. Soc., 254, 177

\bibitem[\protect\citename{Damour \& Taylor }{1992}]{dt92}
Damour~T., Taylor~J.~H., 1992, Phys. Rev. D, 45, 1840

\bibitem[\protect\citename{Hejlesen }{1987}]{hej87}
Hejlesen~P.~M., 1987, A\&AS, 69, 251

\bibitem[\protect\citename{Johnston {\rm et~al. }}{1992}]{jlm+92}
Johnston~S., Lyne~A.~G., Manchester~R.~N., Kniffen~D.~A., D'Amico~N., Lim~J.,
  Ashworth~M., 1992, Mon. Not. R. astr. Soc., 255, 401

\bibitem[\protect\citename{Kaspi {\rm et~al. }}{1996}]{kbm+96}
Kaspi~V.~M., Bailes~M., Manchester~R.~N., Stappers~B.~W., Bell~J.~F., 1996,
  Nature, 381, 584

\bibitem[\protect\citename{Lai, Bildsten \& Kaspi }{1995}]{lbk95}
Lai~D., Bildsten~L., Kaspi~V.~M., 1995, Astrophys. J., 452, 819

\bibitem[\protect\citename{Lorimer {\rm et~al. }}{1993}]{lbdh93}
Lorimer~D.~R., Bailes~M., Dewey~R.~J., Harrison~P.~A., 1993, Mon. Not. R. astr.
  Soc., 263, 403

\bibitem[\protect\citename{Lyne \& Graham-Smith }{1998}]{ls98}
Lyne~A.~G., Graham-Smith~F., 1998, Pulsar Astronomy.
\newblock Cambridge University Press

\bibitem[\protect\citename{Lyne {\rm et~al. }}{1998}]{lml+98}
Lyne~A.~G. {\rm et~al.}, 1998, Mon. Not. R. astr. Soc., 295, 743

\bibitem[\protect\citename{Manchester {\rm et~al. }}{1996}]{mld+96}
Manchester~R.~N. {\rm et~al.}, 1996, Mon. Not. R. astr. Soc., 279, 1235

\bibitem[\protect\citename{Nice, Sayer \& Taylor }{1996}]{nst96}
Nice~D.~J., Sayer~R.~W., Taylor~J.~H., 1996, Astrophys. J. Lett., 466, L87

\bibitem[\protect\citename{Peters }{1964}]{pet64}
Peters~P.~C., 1964, Phys. Rev., 136, 1224

\bibitem[\protect\citename{Phinney \& Kulkarni }{1994}]{pk94}
Phinney~E.~S., Kulkarni~S.~R., 1994, Ann. Rev. Astr. Ap., 32, 591

\bibitem[\protect\citename{Standish }{1982}]{sta82}
Standish~E.~M., 1982, Astr. Astrophys., 114, 297

\bibitem[\protect\citename{Staveley-Smith {\rm et~al. }}{1996}]{swb+96}
Staveley-Smith~L. {\rm et~al.}, 1996, Proc. Astr. Soc. Aust., 13, 243

\bibitem[\protect\citename{Taylor \& Cordes }{1993}]{tc93}
Taylor~J.~H., Cordes~J.~M., 1993, Astrophys. J., 411, 674

\bibitem[\protect\citename{Taylor {\rm et~al. }}{1995}]{tmlc95}
Taylor~J.~H., Manchester~R.~N., Lyne~A.~G., Camilo~F. 1995.
\newblock Unpublished (available at ftp://pulsar.princeton.edu/pub/catalog)

\bibitem[\protect\citename{Thorsett \& Chakrabarty }{1999}]{tc99}
Thorsett~S.~E., Chakrabarty~D., 1999, Astrophys. J., 512, 288

\bibitem[\protect\citename{van Kerkwijk \& Kulkarni }{1999}]{vk99}
van Kerkwijk~M., Kulkarni~S.~R., 1999, Astrophys. J., 516, L25

\bibitem[\protect\citename{Verbunt }{1993}]{ver93}
Verbunt~F., 1993, Ann. Rev. Astr. Ap., 31, 93

\bibitem[\protect\citename{Wex }{1998}]{wex98}
Wex~N., 1998, Mon. Not. R. astr. Soc., 298, 67

\bibitem[\protect\citename{Will }{1993}]{wil93}
Will~C.~M., 1993, Theory and Experiment in Gravitational Physics.
\newblock Cambridge University Press, Cambridge

\end{thebibliography}

\end{document}